\documentclass[11pt]{article} 
\usepackage{hyperref} 
\pdfoutput=1
\begin{document} 

\title{Hexane Air Combustion Behavior\\ Schlieren Visualization}
\author{Philipp A. Boettcher, Brian Ventura, Joseph E. Shepherd \\ 
\\\vspace{6pt} Explosion Dynamics Laboratory, Graduate Aerospace Laboratories \\California Institute of Technology, Pasadena, CA 91125, USA}

\maketitle

\begin{abstract} Hot surface ignition and subsequent flame propagation of premixed n-hexane air mixtures are shown in this fluid dynamics video. High speed schlieren photography revealed 3 distinct behaviors of ignition and propagation as a function of mixture composition and initial pressure. \end{abstract}


\section{Introduction}
Experiments were performed in a closed 2 liter vessel, filled to a precision of 0.1 Torr using the method of partial pressures. The video shows schlieren images from 3 fuel rich mixtures of n-hexane in air, 3.4\%, 4.74\%, and 6.23\%, which elucidate the three different behaviors: single flame propagation, multiple flames, and puffing flames. 

The first mode involves a single flame propagating until it reaches the vessel walls as shown in the first sequence.  In the second mode, two to three flames ignite sequentially, as shown in the second schlieren video.  The final mode corresponds to a continuously puffing flame as seen throughout the duration of the video (about six seconds), of which a portion is shown here. All three experiments were performed at 101kPa initial pressure and the mixtures ignite when the glow plug reaches around 900 K.

This work has been submitted to the Eighth International Symposium on Hazards, Prevention, and Mitigation of Industrial Explosions and is currently under review.


%

\end{document}